# A two-coil mutual inductance technique to study matching effect in disordered NbN thin films

Sanjeev Kumar[a], Chandan Kumar[a], John Jesudasan[b], Vivas Bagwe[b], Pratap Raychaudhuri[b] and Sangita Bose[a*]

[a] Center for Excellence in Basic Sciences, University of Mumbai, Vidhyanagari Campus, Mumbai-400098, India.
[b] Tata Institute of Fundamental Research, Homi Bhabha Road, Colaba, Mumbai 400005.

*Abstract: Although matching effects in superconducting anti-dot arrays have been studied extensively through magneto-resistance oscillations, these investigations have been restricted to a very narrow temperature window close to the superconducting transition. Here we report a "two coil" mutual inductance technique, which allows the study of this phenomenon deep in the superconducting state, through a direct measurement of the magnetic field variation of the shielding response. We demonstrate how this technique can be used to resolve outstanding issues on the origin of matching effects in superconducting thin films with periodic array of holes grown on anodized alumina membranes.*

---

[*] Email: sangita@cbs.ac.in

## Letter

In recent years, the matching effect observed in superconducting films with periodic array of holes, has been a subject of considerable theoretical [1,2,3,4] and experimental attention [5,6,7,8,9,10,11,12,13,14,15,16]. In magnetic fields, matching effect manifests as a periodic variation of superconducting transition temperature ($T_c$) and critical current ($I_c$) with a period commensurate with integer number of flux quantum ($\Phi_0$) passing through each unit cell. This has been explained through two distinct but not necessarily mutually exclusive mechanisms. The first mechanism is related to Little-Parks effect in a single superconducting loop, where $T_c$ mimics the periodic variations in the supercurrent resulting from flux quantization [17]. A film with periodic array of holes can be visualized as a periodic network of side sharing loops, provided the width of the superconducting region is smaller than the Ginzburg-Landau coherence length ($\xi$). It has been theoretically shown that in such networks quantum interference (QI) effects would also give rise to $T_c$ oscillations similar to single-loop Little Parks [3], though additional weaker structures could appear at non-integral rational fractions of flux filling [4]. The second mechanism is related to vortex pinning. It has been argued that the pinning of the vortex lattice is enhanced through the commensurate pinning (CP) when each hole contains an integer number of vortices, thereby giving rise to periodic variation in $I_c$. So far there is little consensus on the relative importance of these two effects [5,6,7,15]. The main reason is that study of the matching effect through the magnetic field variation of R(H) or $I_c$(H) is restricted to temperatures very close to $T_c$. Therefore, while many attribute the matching effect to CP at these temperatures [5,8,9,11], the role of Little-Parks like QI effect cannot be ruled out since $\xi \propto (1 - T/T_c)^{-1/2}$ can become larger than inter-hole separation close to $T_c$ [7,12,16,18]. The focus of the present study is to demonstrate a new two-coil mutual inductance technique [19,20] which probes the matching effect through the magnetic shielding response of the sample at temperatures which is inaccessible in conventional magneto-resistance measurements.

Superconducting NbN thin films were deposited through reactive DC magnetron sputtering on free-standing nanoporous anodic alumina membranes (AAM) obtained from Synkera Technologies Inc. This method to grow superconducting anti-dot arrays has been used by several groups [7,10,12]. AAM with a nominal hole size of 18 nm was used with pore period of 44 nm and the holes were ordered in a hexagonal pattern. NbN thin films were deposited on the AAM substrates by sputtering Nb in Ar/$N_2$ gas mixture keeping the substrate at $600^0$ C. It has been shown that for NbN, $T_c$ is sensitive to the degree of disorder in the form of Nb vacancies in the crystalline lattice, which can be controlled by tuning the deposition conditions [21]. For this study we concentrated on two films with two different levels of disorder: A low disorder film with $T_c$ ~ 12.1 K and a high disorder one with $T_c$ ~ 4.2 K. Scanning electron microscopy (SEM) images of the AAM before and after the deposition reveals an expansion of both pore size (25 nm) and pore period (53 nm). To prevent any damage to the delicate membrane after deposition, the AAM was transferred and glued on a silicon substrate. The thickness of the films was between 25-30 nm. Figure 1 (a) shows the SEM image of a typical film grown on the template.

Figure 1(b) shows the schematics of the "two-coil" apparatus used to measure the screening response of the nanoporous NbN films. The films are sandwiched between two miniature coils: (i) A quadrupolar primary coil having 28 turns with the upper half wound in one direction and the lower half wound in the opposite direction and (ii) a dipolar secondary coil consisting of 112 turns wound in 4 layers. The peak field produced by the primary coil is 10 mG/

mA. The quadrupole configuration of the primary coil ensures a fast radial decay of the magnetic field such that edge effects can be minimized. For our films and coil configuration the supercurrent at the edge of the film is two orders of magnitude smaller than the peak supercurrent at the centre. The whole assembly is made of insulating materials such as Macor and Delrin to avoid spurious signals from eddy currents. A small excitation current, $I$ (0.5 mA, 60 kHz), is passed through the primary and the *in-phase* and *out-of-phase* voltage (V' and V") induced in the secondary is measured using a lock-in amplifier. The real and imaginary part of the mutual inductance (M' and M") corresponding to the inductive and dissipative part, is given by $M'^{(")}=V'^{(")}/(2I\pi\nu)$, where $\nu$ is the frequency of the excitation current. The measurements are performed either in a conventional $^4$He flow cryostat fitted with an 8 T superconducting magnet down to 2 K or in a $^3$He cryostat fitted with a 6 T superconducting magnet down to 300 mK. The $^3$He cryostat is designed to have the sample and coil assembly immersed in $^3$He inside the pot to avoid temperature gradients.

Figure 1 (c) & (e) show the temperature variation of M' and M" for the two NbN films used in this study. Figures 1(d) & (f) show the corresponding temperature variation of the resistance (R-T) for the two films. M'-T curves show a clean superconducting transition. $T_c$ is determined from the intersection of two tangents drawn above and below the transition. It is interesting to note that R-T on the same films show a two step transition which is more pronounced for the film with lower $T_c$. For both films $T_c$ determined from M'-T coincides with the lower temperature where the resistance goes below the limit of our measurement. We do not ascribe this with sample in-homogeneity resulting from two different phases for two reasons. First, NbN films deposited in the same run on MgO substrate shows a clean transition in the R-T.

Secondly, a double transition arising from two phases with different $T_c$ should reflect as a double peak in M''(T) (Figure 1 (c) and (e)) which are not observed here. We believe that the two transitions are related with the establishment of the global phase coherence in the superconducting network. It has been demonstrated that superconducting films deposited on AAM have a significant thickness variation determined by the local slope [22] on the AAM. The system thus behaves like a network of weakly coupled superconducting islands separated by regions which are either metallic or superconducting with a lower $T_c$. At the upper transition the superconducting order sets in on the islands which remain phase incoherent with respect to each other. At the lower transition, the global phase coherence sets in between these islands giving rise to a zero resistance state. Since a global shielding current is established only when the resistance is zero, the shielding response M' is sensitive only to the lower transition. It is worthwhile to note that similar two transitions are observed in mesoscopic normal-metal superconducting arrays [23]. Since the phase stiffness of the superfluid decreases with increase in disorder [24] in NbN thin films, the double transition manifests itself more clearly in the resistive transitions for the more dis-ordered sample ($T_c$ = 4.2 K) compared to the less dis-ordered sample ($T_c$ = 12.1 K).

To show how matching effects can be probed with this technique M' and M'' was measured as a function of magnetic field at different temperatures below $T_c$. Figure 2(a-d) show M' - B and M'' - B for the two films studied. We observe clear oscillations with field in both the films. The period of oscillation, 8.9 kG, is in close agreement with the value expected for the increment of one flux quantum per unit cell in the hexagonal lattice on AAM, given by the formula, $B_M = \Phi_0/A$ where $A = 2\Phi_0/\sqrt{3}d^2$ and d is separation between the pores. The matching effect are also observed in the field variation of M"(B). In this case however, we observe a gradual evolution from a peak to a dip in M"(B) at the matching field as the temperature is increased towards $T_c$. To understand this we note that the matching effect also manifests as an oscillation in $T_c$, with $T_c$ being maximum at matching fields. This is shown in figure 3(a)-(b) where we plot the variation of M'-T and M"-T at the matching field and midway between two matching fields. Using the same criterion as the one used for zero field, we plot the variation in $T_c(H)$ in Fig. 3(c) and (d), showing an oscillatory variation of $T_c(H)$. Similarly, for M"-T the dissipative peak temperature within one period varies between a maximum, $T_p^{Max}$, at the matching field and a minimum, $T_p^{Min}$ at a field in between two matching fields. It is easy to see that when $T < T_p^{Min}$ the matching effect in M" manifests as a peak whereas when $T < T_p^{Max}$ it manifests as a dip. When the temperature $T_p^{Max} > T > T_p^{Min}$ we expect a more complicated pattern with split peak/dip structure consistent with our observation.

The importance of our technique can be gauged from the temperature down to which oscillations are observed in M'-B. In earlier reports the temperature window for observing matching effects was restricted to 0.9-0.8$T_c$ [5,11,13,14]. In this work we see that for the sample with $T_c$ ~ 12.1 K, the matching effect persist down to a reduced temperature, t = T/$T_c$ ~ 0.6. For the sample with $T_c$ ~ 4.2 K matching effects in M'(B) are observed down to the lowest temperature of our measurement, 0.38 K, corresponding to t ~ 0.09. It is worth noting that the resistance at these reduced temperatures (t) is zero (see figure 1(d) and 1(f)) hence magneto-resistance oscillations is not expected to persist in these regimes. Thus, our technique provides a large dynamic range to probe matching effects in the superconducting state.

Next we explore the origin of the matching effect to such low temperatures. The matching effect in lithographically fabricated superconducting wire networks, where the wire diameter is less than $\xi$, has normally been attributed to QI effects [25, 26]. Recently, the matching effect in Nb films deposited on AAM has also been attributed solely to QI effects [12]. For NbN, $\xi \sim 4.5$ nm for $T_c \sim 12.1$ K sample and $\xi \sim 6.5$ nm for $T_c \sim 4.2$ K sample [27]. On the other hand, the width of the superconductor between the holes is $w \sim 27$ nm ($w = d - a$, where $d$ is the separation between the pores and $a$ is the pore diameter). Thus our samples will behave like a wire network only in a narrow temperature range above t ~ 0.96. We can also compare the amplitude of the $T_c$ oscillation expected in a wire network arising from QI as given by [7,28,29]:

$$\frac{\Delta T_c}{T_{c0}} = -0.73 \left(\frac{\xi_0}{d/2}\right)^2 \left[\left(n - \frac{\varphi}{\varphi_0}\right)^2\right]$$

where, n, is an integer, $\phi$ is the flux corresponding to the applied magnetic field and $\phi_0$ is the flux corresponding to the first matching field. $\Delta T_c$ using the above formula ranges between 7-50 mK as also observed in Nb films with anti-dot arrays [29]. However our observed $\Delta T_c$ is almost an order of magnitude larger ruling out QI being the dominant mechanism. On the other hand, CP can also lead to $T_c$ oscillations. In CP reduced pinning at non-integer flux filling allows flux flow of the vortices in a driving field to happen at a lower temperature when the field is intermediate between two matching fields than when the field is at the matching field. Since the onset of the flux flow results in the destruction of the zero resistance state, the periodic variation in the flux flow onset temperature would manifest as a periodic variation of $T_c$ determined from M′-T measurements. We believe that the large oscillation in $T_c$ observed in our sample is a manifestation of CP.

Further evidence of CP is obtained by comparing the temperature range over which the matching effect is observed in the two films. The flux tube threading a hole in the AAM will experience two pinning forces: The pinning force due to the confinement potential of the hole, and the net pinning force exerted on a vortex from its interaction with all other neighboring vortices. The role of CP will be significant when the latter is a significant fraction of the former. The repulsive force between two vortices at a distance R from each other is proportional to $\lambda/R$ [30] where $\lambda$ is the penetration depth. For NbN films $\lambda$ increases rapidly with increase in disorder, from $\lambda(0) \sim 400$ nm for $T_c \sim 12.1$ K to $\lambda(0) \sim 2$ μm for $T_c \sim 4.2$ K. Therefore one would expect the interaction to be about 5 times stronger for the sample with $T_c \sim 4.2$ K. At the same time since condensation energy of the superconductor with lower $T_c$ is expected to be smaller, the confining potential will be weaker. Therefore one would expect the matching effect due to CP to persist down to a lower t for the sample with lower $T_c$, in agreement with our observation.

In conclusion we have shown how measurements of the mutual inductance using the two-coil technique can be used to study matching effects in an anti-dot array of a superconducting film. We have demonstrated that this technique is more powerful and versatile over the more commonly used magneto-resistance measurements because the phenomena can be probed deep in the superconducting state thereby giving us a handle to conclusively track the origin of the matching effect in situations where this effect is not restricted only close to $T_c$. We have also demonstrated that a large periodic variation in $T_c$ can originate from CP. However, a quantitative theory to calculate the magnitude of this effect from CP is currently lacking and its formulation would definitely enhance our understanding of matching effects in nano-perforated films.

**Figure Captions:**

**Figure 1(Colour Online): (a) The Scanning Electron Micrograph (SEM) of NbN thin film deposited on nanoporous anodic alumina membrane (AAM). After deposition at $600^0C$ the pore diameter is found to be 25 nm and pore to pore separation is 53 nm. (b) Schematic Diagram of the "two-coil" apparatus. (c) & (e) Temperature variation of the real (M') and imaginary (M") part of the mutual inductance for the NbN thin films grown on AAM showing a $T_c$ of 12.1 K and 4.2 K respectively. (d) & (f) Temperature variation of the resistance for the corresponding samples; The R-T for the sample with $T_c \sim 4.2$ K is plotted in log-scale for clarity.**

**Figure 2 (Colour Online): (a) & (b) M' as a function of magnetic field for the films with $T_c$ of 12.1 K and 4.2 K at different temperatures below $T_c$. (c) & (d) M" as a function of magnetic field for the same films at different temperatures.**

**Figure 3 (Colour Online): Representative temperature variation of (a) M' and (b) M" at matching and midway between matching fields for the NbN film grown on AAM with $T_c \sim$ 12.1 K. (c) & (d) Red curves shows the $T_c$ as a function of magnetic field obtained from M' - T measurements for the two films with $T_c \sim$ 12.1 and 4.2 K respectively. Black curves show the matching effect from the magnetic field variation of M' at a temperature close to $T_c$. $T_c$ is maximum at the matching fields.**

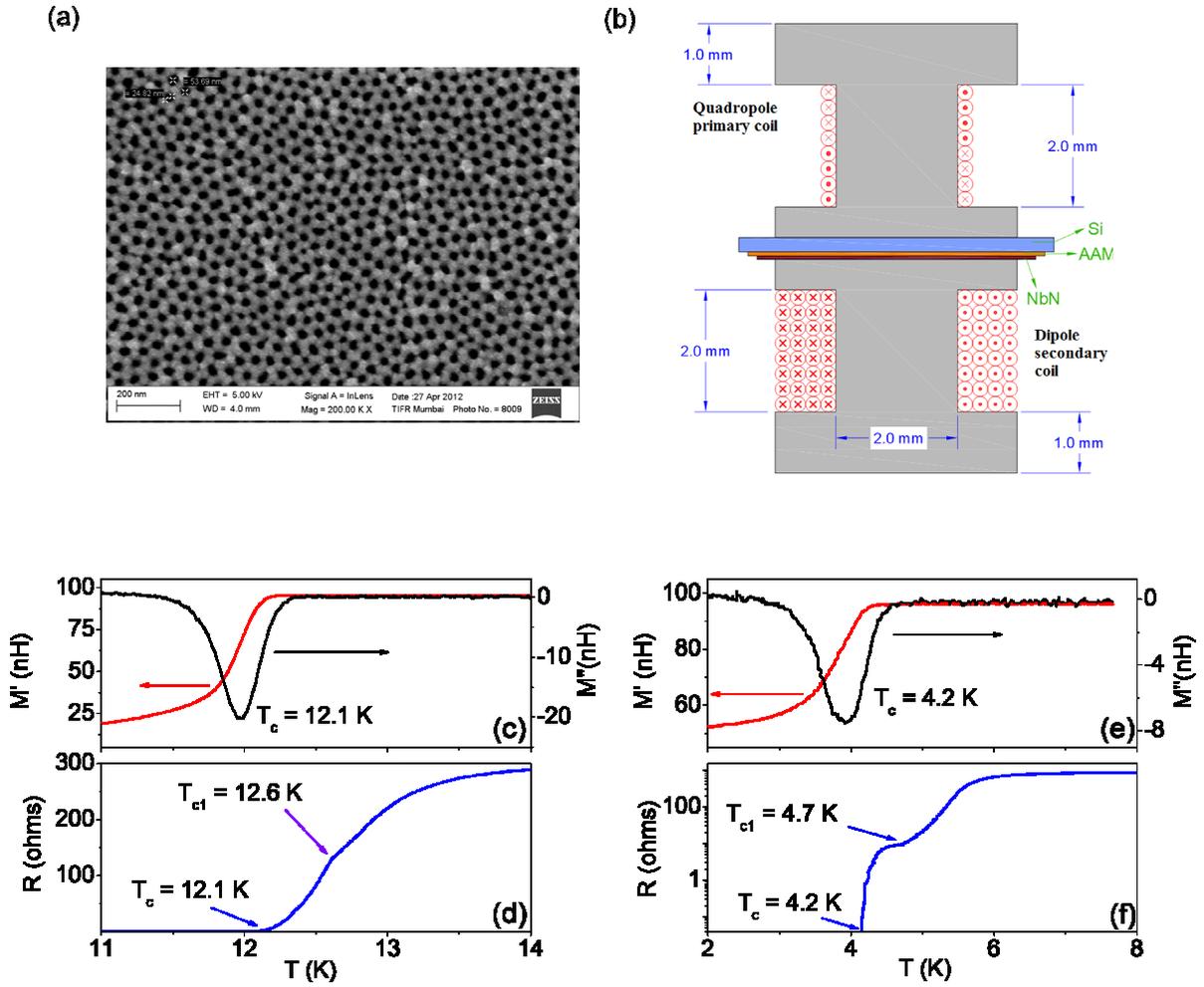

Figure 1

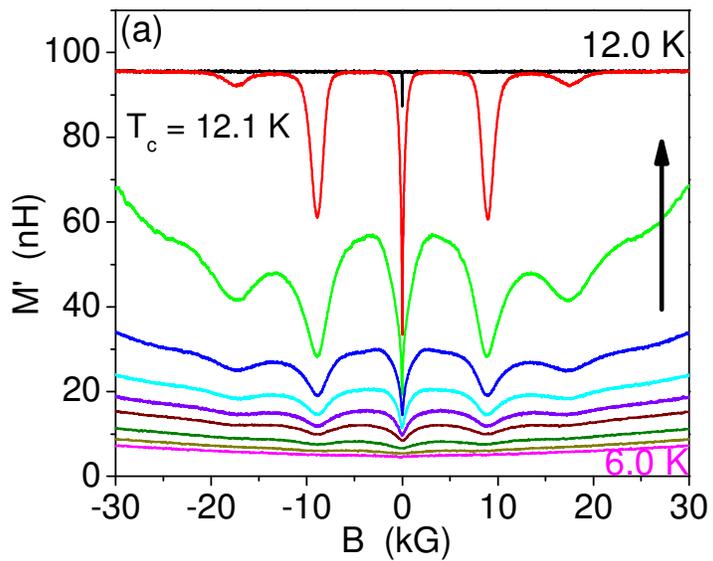
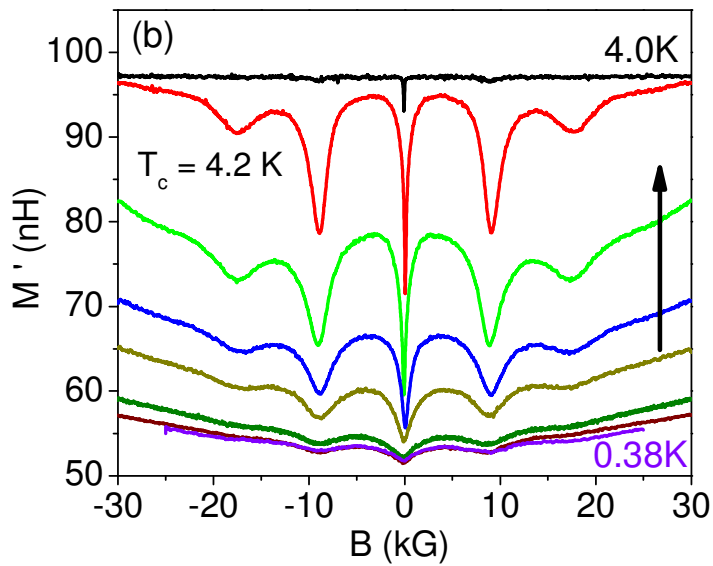
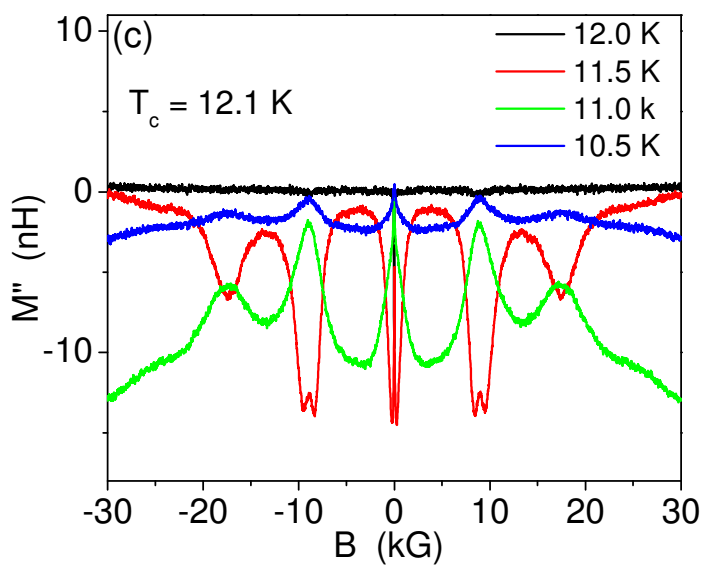
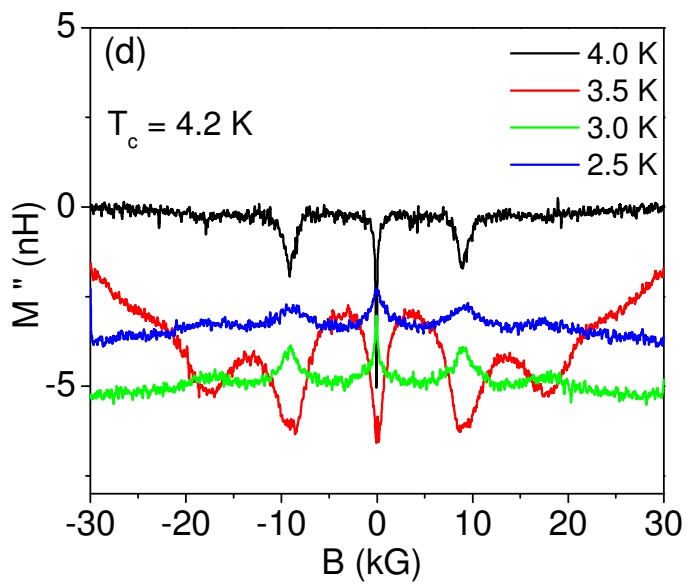

Figure 2

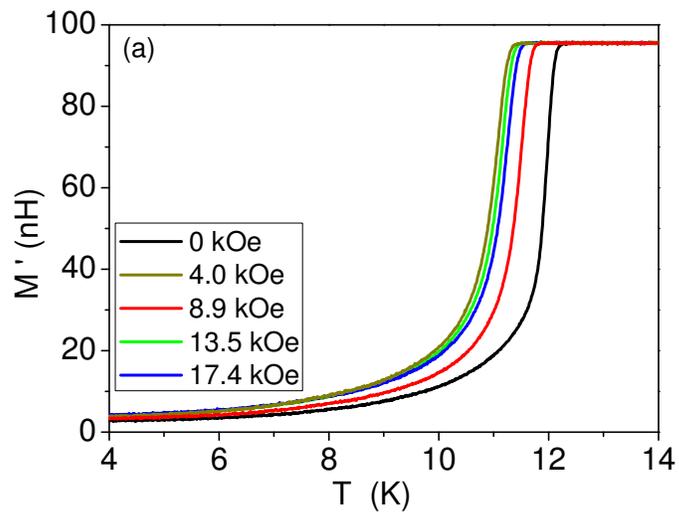
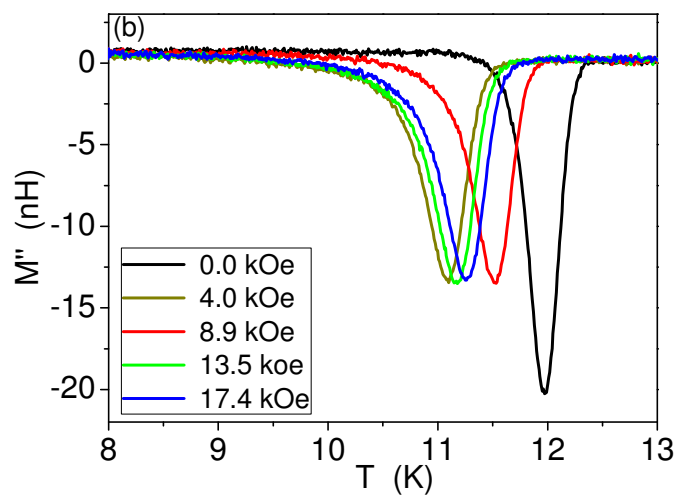
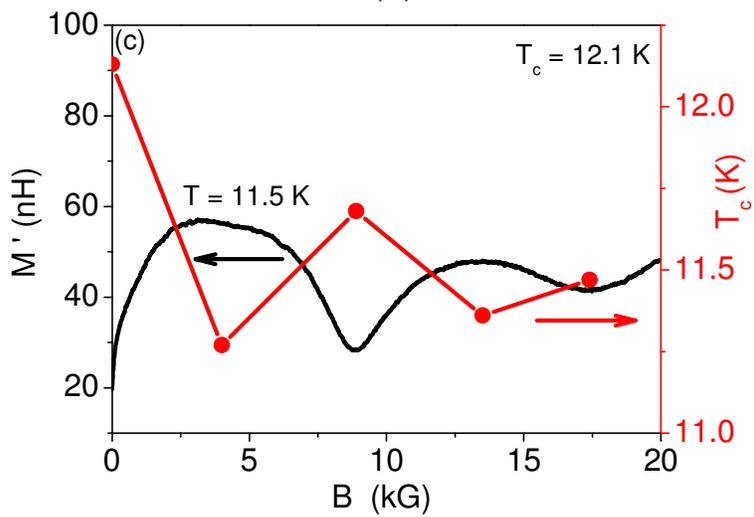
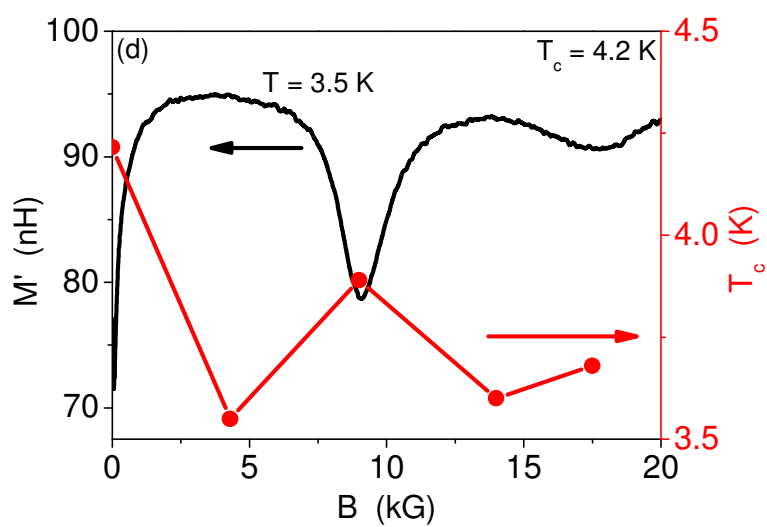

Figure 3

15